# Comment on: Increasing Exclusion: The Pauli Exclusion Principle and Energy Conservation for Bound Fermions are Mutually Exclusive
arXiv:physics/0609190v4 [physics.gen-ph] by Jonathan Phillips

Joshua B. Halpern, Department of Chemistry, Howard University, Washington, DC 20059

**Abstract:** Phillips incorrectly analyzes the ionization of He to conclude that the Pauli principle and conservation of energy are mutually exclusive. His error arises from neglect of electron repulsion, improper choices of energy zeros and other trivial errors.

Phillips has submitted four drafts of a paper to arXiv asserting the Pauli exclusion principle and energy conservation are mutually exclusive for bound fermions [1]. Previously I commented on the basic error Phillips makes [2]. Here a few new issues are treated and the original comment is simplified. Phillips' astounding claims are first quoted, and then it is show that they arise from an elementary misunderstanding. There is no contradiction between the Pauli exclusion principle and energy conservation, nor between quantum mechanics and either of these.

Phillips first considers a model he calls descriptive quantum mechanics (DQM). He purports to analyze the ionization of He. On page 6 of the fourth draft of Ref. 1 he states that:

*A second aspect of the model of relevance is the postulate that for all two electron atoms in their ground state, the two electrons are identical in all respects, except for their spin direction/spin quantum number. In particular, the two electrons have identical energy. For helium the energy of both electrons in the DQM model is approximately -24.5 eV [ 3], because, that is the energy required for ionization. The energy assignment of - 24.5 is the only rational interpretation of the 'energy level diagram' repeatedly employed in descriptions of DQM. In the DQM model the magnitude of the ionization energy and the magnitude of the energy of the electron that is being ionized are equal, just as it is the case for one electron systems. That is, the value -24.5 eV does not come necessarily from a direct quantitative theory (and DQM as defined here is not quantitative), but rather from objective scientific observation (14) of the ionization process*

*A third aspect of DQM of relevance is that after/during ionization the remaining electron in $He^+$ 'falls' to a lower energy state. The measured value of ionization for the electron in $He^+$ is -54.4 eV, a value in close agreement with the ionization calculated using Schrodinger's equation for the ground state of a one electron atom with a two proton nucleus. Although it is not generally explicitly stated, in virtually any standard text (15-17), it is an inescapable implication of the model, and the only interpretation consistent with energy level diagrams (see Figure 1) and the PEP. Thus, the electron that remains after He is ionized 'falls', according to this theory, from approximately -24.7, to -54.4 eV*

And on page 8 he essentially repeats this argument

*1. During/after ionization the electron that is not ionized falls in energy from -24.5 to - 54.4 eV. Where do the nearly 30 eV lost by the remaining bound electron in this process go? Since chemical reactions are basically the movement of electrons between states, this conundrum can be readily written in the symbolism of chemical reactions:*

*He (-49.0 eV) + hν (+24.5 eV) → He+(-54.4 eV) + e- (0 eV); $\Delta H_{rxn}$= -29.9 eV (1)*

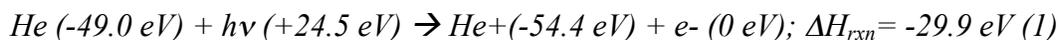

*Where the energy of the species, assuming the Pauli Exclusion Principle is correct, are provided. (Incidentally, the identification of the measured ionization energy, 24.5 eV for helium, and 'energy level' are universal for this form of quantum.) The nearly 30 eV unaccounted for is an*

*enormous energy. In contrast, the re-organization of hydrogen and oxygen to make water produces about 1.2 eV /H atom, and this is associated with a lot of sensible heat. (One plank of the U.S. national energy plan is to capture this very significant energy directly as electricity.) Moreover, it is clear the 1.2 eV is simply the net released by electrons moving from one set of stable orbitals to a new set of orbitals of lower energy that become available for occupancy upon the formation of water molecules. In precisely the same fashion the 29.9 eV of energy released (according to DQM), per Equation 1, represents the net change in the stable energy levels of the electrons in helium during ionization.*

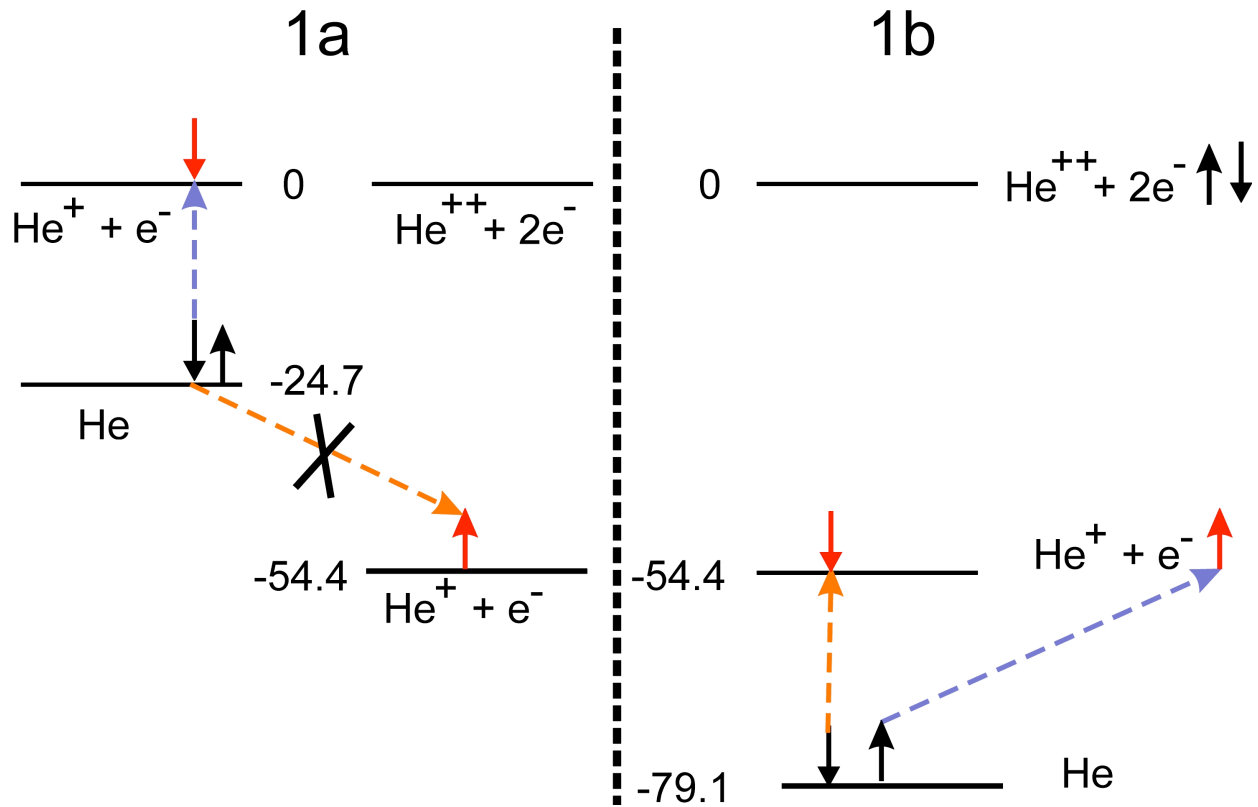

Figure 1a sketches Phillips' argument (see Fig 2a in Reference 1); Fig. 1b shows the actual situation. Electrons are indicated by small solid arrows. Fig. 1b shows that neutral He is ionized by a 24.7 eV photon. This separates one of the electrons from the remaining ion (blue dashed lines) leaving singly ionized $He^+$ whose electron is bound by 54.4 eV. The total energy needed to doubly ionize neutral He is 79.1 eV [4].

Phillips' fatal error is claiming that the ionization process only affects the electron that is removed from the neutral He leaving the energy of the other electron unchanged. Ionizing one of the electrons overcomes the energy of attraction between the nucleus and that electron, but also decreases the repulsion between the two electrons to zero from its positive value. It is this net change which results in a first ionization energy of 24.7 eV. Phillips ignores the repulsion between the two electrons. In his picture, the removal of one electron has no effect on the energy of the other. Thus, the unionized electron in Phillips' model remains at -24.7 eV. Since $He^+$ lies 54.4 eV below $He^{++}$ the ~30 eV difference is claimed as an inconsistency between energy conservation and the exclusion principle. Another way of looking at this is that Phillips chooses $He^+ + e^-$ as the zero of energy for the first ionization (Fig. 1a) and the zero for the second ionization of $He^+$ as $He^{++} + e^-$. He uses a shifted scale for the second step without accounting for it. Phillips' argument is wrong at every level.

Another significant error is Phillips' claim that since the ionization energy of neutral He is 24.7 eV, DQM requires the energy of each electron to be -24.7 eV and the total energy of neutral He to be -49.4 eV. The choice of zero for the enthalpy of formation is arbitrary. A chemist would set the enthalpy of formation of neutral He to zero, in which case singly ionized $He^+$ lies 24.7 eV and $He^{++}$ 79.1 eV above neutral He. A physicist would choose doubly ionized $He^{++}$ as the enthalpy zero, in which case $He^+$ and neutral He lie respectively 54.4 ev and 79.1 eV lower. This has the advantage that the enthalpy zero also corresponds to zero energy for at rest particles. Each electron would on average lie 39.55 eV below doubly ionized $He^{++}$. Correctly written, Eq. (1) above from Phillips should read

He(2x-39.55 eV = -76.1 eV) + hν(+24.7 eV) → $He^+$(-54.4 eV) + e- (0 eV); $\Delta H_{rxn}$= -0 eV (2)

Thus, there is no conflict between the Pauli exclusion principle and conservation of energy in He, or for that matter any multi-electron atom or ion.

Later in the fourth revision, Phillips accepts that quantum mechanics sets an average energy of -39.55 eV for each of the electrons in neutral helium. He calls that the standard quantitative quantum mechanics (SQQM) picture (See his Fig. 2), but the core error remains, ignoring the fact that ionizing He is a balance between pulling an electron away from the nucleus while the electron-electron repulsion is simultaneously reduced to zero. As in the first case this results in an erroneous claim of a conflict between energy conservation and the exclusion principle.

This comment is not concerned with criticisms of the semi-classical theory that Phillips offers which itself has multiple flaws, nor his naïve and incorrect interpretations of quantum mechanics. It establishes that Phillips' justification for offering such a theory, the supposed irreconcilability of quantum mechanics, the Pauli principle and conservation of energy, is without basis.

1. J. Phillips, "Increasing Exclusion: The Pauli Exclusion Principle and Energy Conservation for Bound Fermions are Mutually Exclusive, arXiv:physics/0609190v4 [physics.gen-ph]
2. J. Halpern, "Comment on: Increasing Exclusion: The Pauli Exclusion Principle and Energy Conservation for Bound Fermions are Mutually Exclusive arXiv:physics/0609190v2 [physics.gen-ph] by Jonathan Phillips" arXiv:0709.4065v1 [physics.gen-ph]
3. Obviously 24.7 eV is meant here and throughout
4. Yu. Ralchenko, F.-C. Jou, D.E. Kelleher, A.E. Kramida, A. Musgrove, J. Reader, W.L. Wiese, and K. Olsen **Atomic Spectra Database Version 3.1.0** National Institutes of Standards http://physics.nist.gov/PhysRefData/ASD/index.html